\documentclass{aa}

\usepackage{graphicx}

\begin{document}

\title{\object{IY UMa}: accretion disc evolution after superoutburst}
\author{
V. Stanishev\inst{1}$^\star$
\and
Z. Kraicheva\inst{1}$^\star$
\and
H.M.J. Boffin\inst{2}$^\star$
\and V. 
Genkov\inst{1}\thanks{E-mail: 
{\rm vall@astro.bas.bg} (VS), {\rm zk@astro.bas.bg} (ZK), {\rm
henri.boffin@oma.be} (HMJB), {\rm nao@mail.orbitel.bg} (VG)}
}          

\offprints{Z. Kraicheva}

\institute{Institute of Astronomy, Bulgarian Academy of Sciences, 
           72, Tsarighradsko Shousse Blvd., 1784 Sofia, Bulgaria
   \and
	Royal Observatory of Belgium, 
              Avenue Circulaire 3, B-1180 Brussels, Belgium
 }

\date{Received ; accepted }

\authorrunning{V. Stanishev et al.}

\abstract{
CCD photometry of the newly discovered eclipsing dwarf nova
\object{IY UMa} is used to study the physical properties of 
the accretion disc in the late
decline and quiescence stages. Eclipse mapping analysis shows that in
these stages the accretion disc is cool 
with approximately flat radial brightness temperature distribution
and $T_{\rm BR}\sim5000-5500$\,K. The hot spot is found to lie close along
the stream trajectory at distance $0.36a$  and $0.25a$ from the
disc center in February and March.
The orbital hump strength
decreases by $\sim30$\% in a month. Taking the smallest possible
size of the hot spot, we derive an upper limit for its brightness
temperature, $\sim15900$\,K and $\sim13800$\,K, respectively.
}

\maketitle

\section{Introduction}

An intensive photometric monitoring of the recently discovered dwarf nova
Tmz~V85 (Takamizawa \cite{taka}) was started after January 13, 2000 when
Schmeer (\cite{sch}) reported a new outburst. The star has been classified
as a deep eclipsing SU~UMa type dwarf nova by Uemura et al. (\cite{uema})
and got the designation \object{IY UMa} (Samus \cite{samu}). Using numerous 
long photometric
series covering the whole superoutburst, Patterson
et al. (\cite{pat}) determined the system parameters:
orbital period $P_{\rm orb}=$0\fd0739092, masses
of the two components $M_1=0.68M_{\sun}$ and $M_2=0.10M_{\sun}$, inclination
$i$=85.9$^\circ$ and distance $D$=190~pc. During the superoutburst the star
showed both ``normal'' and ``late'' superhumps with a period of
$\sim$0\fd07583.

Here we report our time-resolved photometric observations of \object{IY~UMa}
started about
the end of the late superhumps stage and continued after the
accretion disc shrank back in quiescence.

\section{Observations and data reduction}

The observations of \object{IY UMa} were carried out at Rozhen Observatory
with the 2.0-m telescope and its attached Photometrics
1024x1024 CCD camera. The star was observed in the Johnson $V$ band for three
nights, with an exposure time of 60~s
on February 6 and March 11, 2000 and
with 90~s on March 30, 2000. In addition, $B$ exposures were regularly taken
during the nights.
In each case, there was a $\sim$15 s dead-time.
In total, seven eclipses were covered: three on February 6 and
March~11 and one on March~30.
After de-biasing and flat-fielding the photometry was done with
the standard DAOPHOT procedures (Stetson \cite{ste}). On February 6 and March~30
the seeing was $\sim$2$^{''}$ and we performed PSF fitting photometry,
which allowed to estimate the magnitudes of the faint NE companion:
$V$=19.5 and $B-V$=0.69. The data on
March~11 were processed using aperture photometry including this
star and subtracting its flux afterwards. For all the observations
the stars ID~9 with $V$=17.633~$\pm$0.037 and
$B-V$=1.145~$\pm$0.027 and ID~10 with $V$=17.844~$\pm$0.016 and
$B-V$=0.620~$\pm$0.136 from the list of Henden (\cite{hen}) served as a
comparison star and check, respectively.
The instrumental $b-v$ colors were transformed to
standard ones in the usual way with a second-order extinction
term $k^{''}_{b-v}$=$-0.035$ included. All the $B$ frames were taken outside
eclipse, therefore the $B-V$ colors are representative for these orbital phases
only. The $B-V$ color was found to be constant with the orbital phase
and averages $\sim$~0.0 and $\sim+$0.22 for February and March
observations, respectively. These values were used to convert the
instrumental $v$ band measurements to standard ones. The
magnitudes were converted to fluxes using the absolute calibration of
the $UBV$ system (Strai\v{z}ys \cite{stra}).

\begin{figure}
\includegraphics*[width=8.8cm]{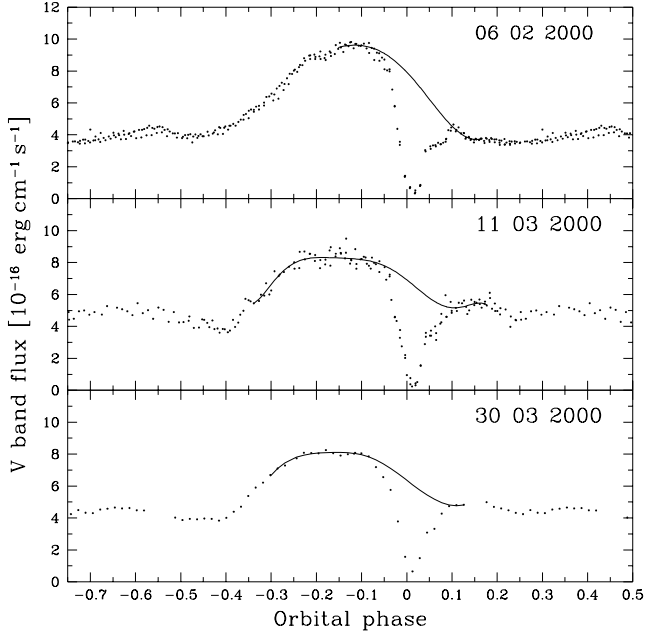}
\caption{
 Photometric light curves, folded with the orbital period
 $P_{\rm orb}=$0\fd0739092. The low 
 order polynomial fits to the out-of-eclipse data are also shown.}
 \label{curves}
\end{figure}

\section{Accretion disc eclipse mapping}

\begin{figure}
\includegraphics*[width=8.8cm]{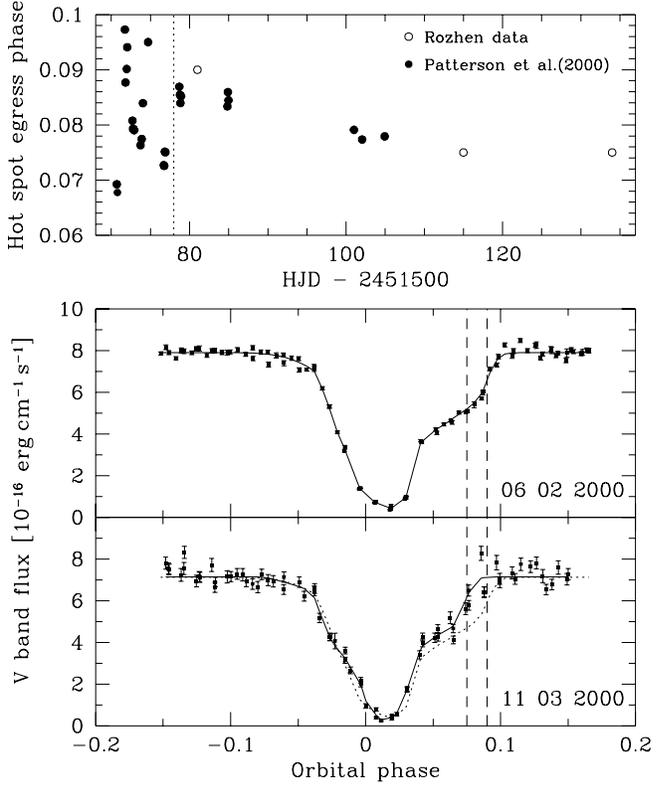}
\caption{ {\it Upper panel:} Hot spot egress phases. The trend
after JD2451578 indicates a decrease of accretion disc radius.
Vertical line marks the end of late superhumps era.
{\it Lower panel:} Light curves of the reconstructed disc images. 
Dotted line is the reconstructed February light curve plotted over 
March data to see the difference clearly.
Vertical dashed lines show the hot spot mid-egress phase.}
\label{egress}
\end{figure}

In Fig.\,\ref{curves} the light curves are shown, folded with the ephemeris
determined by Patterson et al. (\cite{pat}). The analysis of these curves was
performed using the system parameters determined by Patterson et al. (\cite{pat}).
To reconstruct the accretion disc $V$ band brightness distribution we used the
eclipse mapping method (Horne \cite{hor}) and the Maximum Entropy
technique as described by Skilling\&Bryan (\cite{ski}) with a
full azimuthal floating default image (Horne \cite{hor}). Since our data were
obtained with exposure times longer than the WD ingress and egress
($\sim$35~s, Patterson et al. \cite{pat}), we were unable to decompose the
light curves into its individual components: WD, hot spot and
accretion disc.
Thus, the curves had to be analysed without subtracting the WD and hot spot flux.
Instead, we accounted for the variation of the hot spot flux with the
orbital cycle by first fitting the  out-of-eclipse data with a low-order polynomial 
(shown with full lines in
Fig.\,\ref{curves}), then normalizing the data to this fit and finally
rescaling the data to the expected hump flux at zero phase. 
The curve obtained on March~30 was excluded from the analysis because of its long
exposure time, hence its very poor orbital phase resolution. 
The final curves and the fits obtained are shown in
Fig.\,\ref{egress} and it can be seen that the hot spot egress
phase changes from $\sim0.09$ in February to $\sim0.075$ in March.
The hot spot egress phase obtained from
the curve not included in the reconstruction (taken on 30 March)
coincides with that determined from the March~11 data. 
It seems
also that in February the white dwarf and hot spot are eclipsed
simultaneously or at least the hot spot is eclipsed very close after
the white dwarf, while in March the ingress of the two compact
sources seem to be resolved, as was observed for example in \object{Z Cha} and
\object{OY Car} (Wood et al. \cite{wood2, wood3}). In Fig.\,\ref{egress} are also shown the
mid-egress 
phases of the hot spot obtained from our data, compared
with those derived by Patterson et al. (\cite{pat}). It can be seen
that our results are in agreement with a continuous decrease of the
hot spot mid-egress phase
when the system evolved to quiescence.

\begin{figure}
\includegraphics*[width=8.8cm]{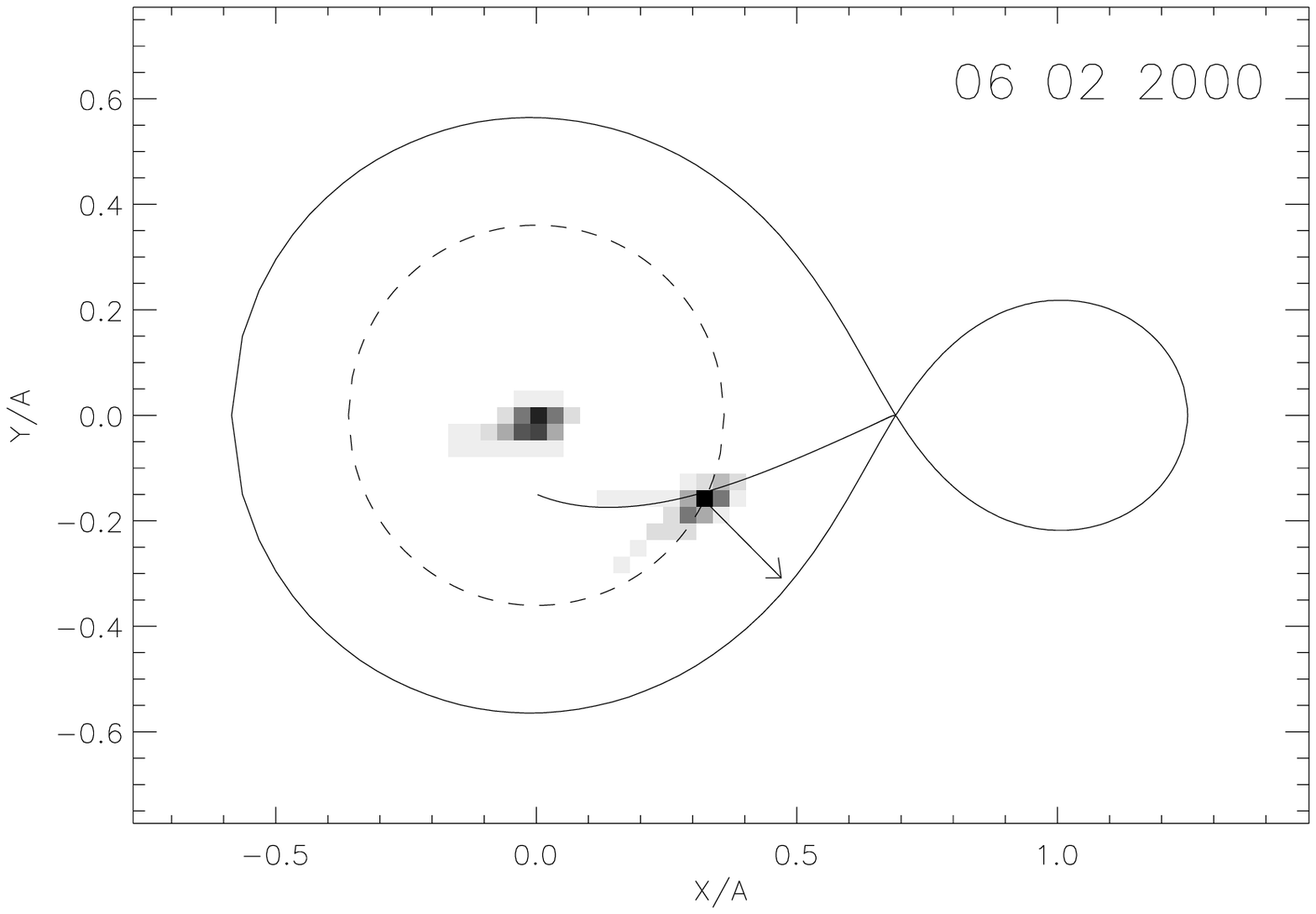} \\
\includegraphics*[width=8.8cm]{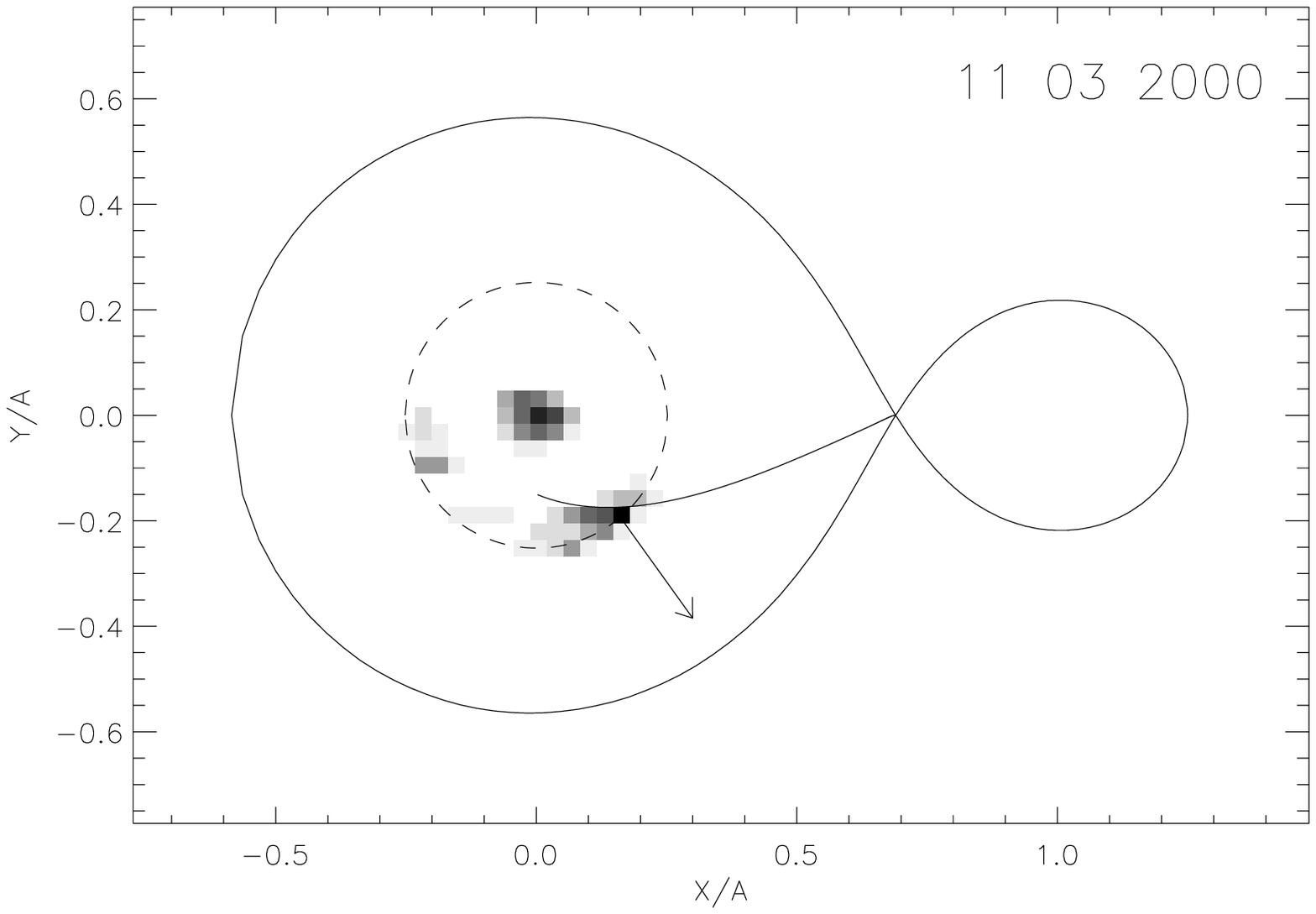}
\caption{Reconstructed brightness distribution of the accretion disc
from February and March data. Arrows mark the direction of the maximal
hot spot visibility.}
\label{mm}
\end{figure}

The calculated brightness distributions are shown in
Fig.\,\ref{mm}. Two compact sources are
found and they are recognised as the white dwarf and the hot spot. The
stream trajectories, calculated by integrating the equations of
motion (Flannery \cite{fla}) by the fourth-order Runge-Kutta method, are also shown
in Fig.\,\ref{mm}. In both maps the
hot spot is found close along the trajectory but at different distance
from the white dwarf, indicating that the accretion disc has
shrank. The radii in February and March, determined by the most
bright pixel in the hot spot, are $r_{\rm d}\simeq 0.36a$ and
$r_{\rm d}\simeq 0.25a$, respectively.
In the disc reconstruction from the March data, something like a third
spot is seen. We believe that this is not a real structure in the disc but is an
artifact from the reconstruction.

\begin{figure}
\includegraphics*[width=8.8cm]{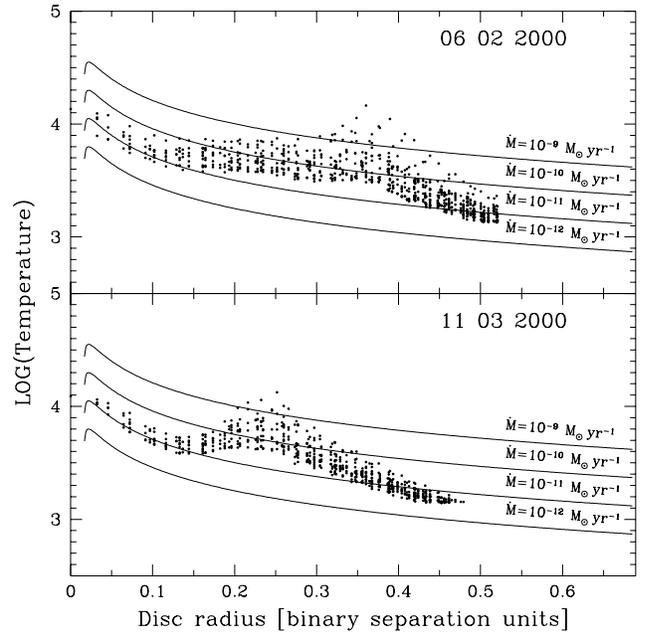}
\caption{Brightness temperature of the disc. Models of steady state
discs for different mass transfer rates are shown also.}
\label{rt}
\end{figure}

Having the brightness distribution calculated we can  estimate the
brightness temperature $T_{\rm BR}$ of the accretion disc assuming black body emission.
This is done by solving for $T_{\rm BR}$ the following equation:
\begin{equation}
f=\frac{\int v(\lambda)B_\lambda(T_{BR})/\lambda\,d\lambda}
{\int v(\lambda)/\lambda\,d\lambda}\,\delta/D^2.
\label{temp}
\end{equation}
Here $f$ stands for the emitted flux, $B_\lambda(T_{\rm BR})$ is the
black body spectral distribution, $v(\lambda)$ is the response of
the $V$ bandpass and $\delta$ is the pixel area corrected for the foreshortening.
In Fig.\,\ref{rt} we show the radial distributions of
the brightness temperature along the accretion disc radius together with the
expected temperatures for steady accretion discs for different mass
transfer rates. The inner part of the disc seems to be
well described as a steady accretion disc model with
$\dot{M}\simeq 10^{-11} M_\odot \,\mbox{yr}^{-1}$. If this is the case
it would be in conflict with other results for dwarf novae
in quiescence where the accretion disc brightness temperature radial
distribution is found
to be much flatter than what is expected for steady discs.
Most probably, however, this is redistributed flux from the white dwarf
as a result of the algorithm.
In the outer part of the disc
reconstructed from the February data, the brightness temperature
distribution becomes almost flat with $T_{\rm BR}\simeq 5500$\,K. In
the March data reconstruction this flat part cannot be seen due to the
smaller disc radius and the redistribution of the white
dwarf and the hot spot fluxes. While $T_{\rm BR}\simeq5500$\,K is a typical
temperature for dwarf novae in quiescence, the observations and
numerical studies show that quiescent accretion discs in these systems
are most probably optically thin. In this case the emission of the
disc significantly deviates from black body law and the above value of
$T_{\rm BR}$ is a very crude estimation of the effective temperature of
the accretion disc.

The prominent humps seen roughly between phases $-0.4$ and $+0.1$
are believed to be produced by the hot spot. The hot spot light is
assumed to be emitted by a circular area perpendicular to the orbital plane,
and the hump is produced by the foreshortening
(Wood et al. \cite{wood2, wood3}). These authors supposed that the size of the
spot is equal to the cross section of the accretion stream (the formula
derived by Cook\&Warner (\cite{cook})) and used Eq.\,\ref{temp} to estimate
$T_{\rm BR}$ of the spot assuming two models of hot spot emission:
\begin{itemize}
\item[i)] the hot spot radiates isotropically from
both its sides with most of the light from nearest to the disc side being
obscured from our sight,
\item[ii)] a two component model in which the hot
spot itself radiates anisotropically from its outer side only and
another isotropic component.
\end{itemize}
Since we cannot decompose our light curves, we used this
method to estimate $T_{\rm BR}$ of the spot corresponding to the
anisotropic part in the second model. The flux $f$ was determined as
the difference between maximal flux and the flux at phases when the 
light of the hot spot is not seen (in March data the fluxes before and
after the hump are significantly different so we took the mean value).
The estimations showed that the strength of the hump decreases roughly
by $\sim30\%$ between February and March, giving 
$T_{\rm BR}\simeq15900$\,K and $T_{\rm BR}\simeq13800$\,K with an error of both 
values $\sim500$\,K. These values are an upper limit for
$T_{\rm BR}$ because the accepted width of the hot spot is a lower limit.
The actual size of the spot is probably much larger as it is produced
by the hypersonic impact between the stream and disc.

\section{Conclusions}

The analysis of the data shows that in spite of the constant system
brightness after February (Fig.~1 in Patterson et al. (\cite{pat}) and our
data) the accretion disc has not reached its equilibrium quiescent
state. Its radius gradually decreases as the system evolves to
quiescence from $r_{\rm d}\simeq0.36a$ in February and $r_{\rm d}\simeq0.25a$ in March. 
This is in agreement with both,
observations and numerical simulations of dwarf nova outbursts (see
review paper of Osaki (\cite{osaki}) and references therein).
In the same time the temperature of the
disc itself seems to remain constant. The light curves
obtained on 11 and 30 March are very similar without any apparent
changes, which suggests that the equilibrium state was most probably
reached in the beginning of March. The eclipse mapping shows a hot
spot, laying along the accretion stream. The strength of the orbital hump
decreases by $\sim30$\%. Assuming that the hump is produced by the hot
spot and that it emits 
anisotropically from its outer side, we estimate an upper limit of the
hot spot brightness temperature $T_{\rm BR}\simeq15900$\,K and
$T_{\rm BR}\simeq13800$\,K in February and March, respectively. Such a
decrease could be produced by an enhanced mass transfer rate during
outburst or/and by a change of the hot spot structure as the angle of
impact and densities of the disc and stream vary with the accretion
disc radius.

\begin{acknowledgements}
V.S, Z.K and V.G are grateful to NFSR for supporting this work with
the project No.~715/97.
\end{acknowledgements}

\end{document}